\documentstyle[11pt,moriond,epsfig]{article}

\def\Journal#1#2#3#4{{#1} {\bf #2}, #3 (#4)}

\def\PLB{{\em Phys. Lett.}  B}
\def\PRL{\em Phys. Rev. Lett.}
\def\PRD{{\em Phys. Rev.} D}

\def\EPC{{\em Eur. Phys. J.} C}

\def\ra{\rightarrow}

\def\be{\begin{equation}}
\def\ee{\end{equation}}
\def\bea{\begin{eqnarray}}
\def\eea{\end{eqnarray}}

\newcommand{\inmath}[1] {\ifmmode#1\else$#1$\fi}
\newcommand{\definmath}[2] {\def#1{\ifmmode#2\else$#2$\fi}}
\def\mrm{\mathrm}
\def\qqbar{\relax\ifmmode{\mathrm{q}\overline{\mathrm{q}}}%
          \else${\mathrm{q}\overline{\mathrm{q}}}$\fi}
\newcommand{\epem}{\mbox{$\mathrm{e^+e^-}$}}

\newcommand{\Zz}{\mbox{${\mathrm{Z}^0}$}}
\newcommand{\WW}{\mbox{$\mathrm{W^+W^-}$}}
\newcommand{\Zg}{\mbox{$\mathrm{Z}^{*}/\gamma^{*}$}}

\newcommand{\qq}{\mbox{$\mathrm{q\overline{q}}$}}
\newcommand{\ff}{\mbox{$\mathrm{f\overline{f}}$}}
\newcommand{\lnu}{\mbox{$\ell\overline{\nu}$}}
\newcommand{\lmnu}{\mbox{$\ell^-\overline{\nu}_{\ell}$}}
\newcommand{\lpnu}{\mbox{${\ell^{\prime}}^+ \nu_{\ell^{\prime}}$}}
\newcommand{\enu}{\mbox{$\mathrm{e\overline{\nu}}$}}
\newcommand{\mnu}{\mbox{$\mu\overline{\nu}$}}
\newcommand{\tnu}{\mbox{$\tau\overline{\nu}$}}
\newcommand{\WWqqqq}{\mbox{\WW$\rightarrow$\qq\qq}}
\newcommand{\WWqqln}{\mbox{\WW$\rightarrow$\qq\lnu}}

\newcommand{\WWlnln}{\mbox{\WW$\rightarrow$\lmnu\lpnu}}
\newcommand{\Wenu}{\mbox{$\epem \rightarrow \mathrm{W}\enu$}}

\newcommand{\Zee}{\mbox{$\epem\rightarrow\Zz\epem$}}

\newcommand{\Mz}{\mbox{$\mathrm{M}_{\mathrm{Z}^0}$}}
\newcommand{\Mw}{\mbox{$\mathrm{M}_{\mathrm{W}}$}}
\newcommand{\Mh}{\mbox{$\mathrm{M}_{\mathrm{H}}$}}

\newcommand{\Opal}{\mbox{O{\sc pal }}}
\newcommand{\Aleph}{\mbox{A{\sc leph }}}
\newcommand{\Delphi}{\mbox{D{\sc elphi }}}
\newcommand{\Lt}{\mbox{L{\sc 3 }}}
\newcommand{\CDF}{\mbox{C{\sc df }}}
\newcommand{\Dz}{\mbox{D{\sc 0 }}}

\newcommand{\roots}{\mbox{$\sqrt{s}$}}

\newcommand{\sinw}{\mbox{$\sin^{2}\theta_{\mathrm{W}}$}}
\newcommand{\mrec}{\mbox{$\mathrm{m_{rec}}$}}

\newcommand{\mrecf}{\mbox{$\mathrm{m_{rec_{1}}}$}}
\newcommand{\mrecs}{\mbox{$\mathrm{m_{rec_{2}}}$}}

\begin{document}
%
% --- title, abscract a francais
%
%%%\vspace*{4cm}
%%%\title{La Mesure de $\mathbf{M_{W}}$ au LEP2}
%%%
%%%\author{ D. Glenzinski }
%%%
%%%\address{Enrico Fermi Institute and Department of Physics, \\
%%%  University of Chicago \\ Chicago, IL 60637, USA }
%%%
%%%\maketitle\abstracts{ En 1997, chaque exp\'erience du LEP ont recuielli 
%%%approximativement $55\:\mrm{pb}^{-1}$ de donn\'ees \`a une \'energie dans le 
%%%centre de masse de $183$~GeV.  Ces donn\'ees procurent un \'echantillon 
%%%$\epem\ra\WW$ et permettent la mesure de la masse du boson W, \Mw. Les 
%%%r\'esultats combin\'es du LEP, incluant les donn\'ees recuiellies \`a
%%%$\roots=161$~et$172$~GeV et utilisant les relations prescritent par le 
%%%Mod\`ele Standard entre la largeur de d\'esint\'egration et la masse du W, est 
%%%$\Mw = 80.35 \pm 0.07 ( \mrm{exp} ) \pm 0.04 ( \mrm{CR} ) \pm 0.03
%%%  ( \mrm{E_{bm}} )$~GeV, o\`u les erreurs correspondent aux incertitudes
%%%exp\'erimentals, aux effets de reconnection de couleur/Bose-Einstein, et \`a
%%%l'incertiture sur l'\'energie du LEP. }

%%%\pagebreak
%
% --- begin document in english
%
\vspace*{4cm}
\title{Measurement of $\mathbf{M_{W}}$ from LEP2}

\author{ D. Glenzinski }

\address{Enrico Fermi Institute and Department of Physics, \\
  University of Chicago \\ Chicago, IL 60637, USA }

\maketitle\abstracts{
In 1997 each LEP experiment collected approximately $55\:\mrm{pb}^{-1}$ of 
data at a center-of-mass energy of $183$~GeV.  These data yield a sample of
candidate $\epem\ra\WW$ events from which the mass of the W boson, \Mw, 
is measured.  The LEP combined result, including data taken at 
$\roots = 161$~ and~$172$~GeV and assuming the Standard Model relation between
the W decay width and mass, is 
$\Mw = 80.35 \pm 0.07 ( \mrm{exp} ) \pm 0.04 ( \mrm{CR} ) \pm 0.03 
  ( \mrm{E_{bm}} )$~GeV, 
where the errors correspond to experimental, colour-reconnection/Bose-Einstein,
and LEP beam energy uncertainties respectively.  }

\section{Introduction}
\label{sec:intro}

\noindent
The success of the Standard Model (SM) over the last two decades should not
obscure the importance of thoroughly investigating the weak interaction.  It
is interesting to consider that 15 years ago, when neutrino scattering 
experiments had measured $\sinw = 0.217 \pm 0.014$, the following 
SM constraints were available~\cite{sirlin}:
\bea
  \Mw ( \mrm{indirect} ) & = & 83.0 \pm 2.8\:\mrm{GeV} \\ 
  \Mz ( \mrm{indirect} ) & = & 93.8 \pm 2.3\:\mrm{GeV}
\eea
{\it Tree level} deviations could be accommodated in those errors!  Today we
have measured~\cite{lepewk} \sinw\ to $0.0002$, \Mz\ to $0.002$~GeV, and 
\Mw\ to $0.07$~GeV --- the success of the SM is so thorough that it can
only be wrong at the quantum loop level, and even then, beyond leading 
order.  Despite this rousing success, it is still necessary to test the SM by 
confronting experimental observations with theoretical predictions as any 
deviations might point to new physics.  As a fundamental parameter of the SM, 
the mass of the W boson, \Mw, is of particular importance.  

Aside from being an important test of the SM in its own right, the direct 
measurement of \Mw\ can be used to set constraints on the mass of the Higgs 
boson, \Mh, by comparison with theoretical predictions involving radiative
corrections sensitive to \Mh.  The constraints imposed using \Mw\ are 
complimentary to the constraints imposed by 
the asymmetry ($\mrm{A_{FB}^{b}}$, $\mrm{A_{FB}^{\ell}}$, $\mrm{A_{LR}}$,...)  
and width ($\mrm{R}_{\ell}$, $\mrm{R_{b}}$, $\mrm{R_{c}}$,...) measurements.  
For example, the very precise asymmetry measurements presently yield the 
tightest constraints on \Mh, but are very sensitive to the uncertainty in 
the hadronic contribution to the photon vacuum polarisation, 
$\Pi^{\gamma\gamma}_{\mrm{had}}$.  In contrast, the 
constraint afforded by a direct measure of \Mw\ is comparably tight but with 
a much smaller sensitivity to $\Pi^{\gamma\gamma}_{\mrm{had}}$,
and is presently dominated by statistical uncertainties~\cite{Mhconst}.

\subsection{WW Production at LEP}
\label{intro-prod}

\noindent
At LEP W bosons are predominantly produced in pairs through the reaction
$\epem\ra\WW$, with each W subsequently decaying either hadronically (\qq),
or leptonically (\lnu, $\ell = e$, $\mu$, or $\tau$).  This yields three 
possible four-fermion final states, hadronic (\WWqqqq), semi-leptonic 
(\WWqqln), and leptonic (\WWlnln), with branching fractions of $45\%$, $44\%$,
and $11\%$ respectively.  The \WW\ production cross-section varies from
$3.6$~pb at $\roots = 161$~GeV to $15.7$~pb at $\roots = 183$~GeV.  These 
can be contrasted with the production cross-sections for the dominant 
backgrounds $\sigma\left( \epem\ra\Zg\ra\qq \right)\approx 110$~pb, 
$\sigma\left( \Zee \right)\approx 2.8$~pb,
$\sigma\left( \epem\ra(\Zg)(\Zg) \right)\approx 0.6$~pb, and 
$\sigma\left( \Wenu \right)\approx 0.6$~pb.  Aside from the 
$\Zg\ra\qq$ process, which falls from $\approx 150$~pb at $\roots = 161$~GeV,
these background cross-sections vary slowly for $\roots < 185$~GeV, when the
$\epem\ra\mrm{ZZ}$ process begins to turn-on.

\subsection{LEP Measurement Techniques}
\label{intro-meas}

\noindent
There are two main methods available for measuring \Mw\ at LEP2.  The first
exploits the fact that the \WW\ production cross-section is particularly
sensitive to \Mw\ for $\roots\approx 2\Mw$. In this threshold (TH) region,
assuming SM couplings and production mechanisms, a measure of the 
production cross-section yields a measure of \Mw .  In early 1996 the four 
LEP experiments collected roughly $10\:\mrm{pb}^{-1}$ of data at 
$\roots=161$~GeV, resulting in a combined determination of the W boson mass of 
$\Mw (\mrm{TH}) = 80.40 \pm 0.20 (\mrm{exp}) \pm 0.03 (\mrm{E_{bm}})$~GeV,
where the errors correspond to experimental and LEP beam energy uncertainties 
respectively~\cite{lepewk,lep161}.

The second method uses the shape of the reconstructed invariant
mass distribution to extract a measure of \Mw.  This method is 
particularly useful for $\roots\geq 170$~GeV where the \WW\ production 
cross-section is larger and phase-space effects on the reconstructed mass
 distribution are smaller.  Each experiment collected roughly 
$10\:\mrm{pb}^{-1}$ at $\roots=172$~GeV in later 1996, and in 1997, roughly 
$55\:\mrm{pb}^{-1}$ at $\roots=183$~GeV.  Since most of the LEP2 data has been
collected at center-of-mass energies well above the \WW\ threshold, the LEP2 
\Mw\ determination is dominated by these direct reconstruction (DR) methods.  
For this reason, the rest of this article will concentrate on the details of 
this method.

\section{Direct Reconstruction of $\mathbf{M_{W}}$}
\label{sec:DR}

\noindent
To measure \Mw\ using direct reconstruction techniques one must
\begin{enumerate}
  \vspace*{-2mm}
  \item Select $\WW\ra\ff\ff$ events.
  \vspace*{-2mm}
  \item Obtain the reconstructed invariant mass,\mrec, for each event.
  \vspace*{-2mm}
  \item Extract a measure of \Mw\ from the \mrec\ distribution.
\end{enumerate}
Each of these steps are discussed in detail in the section below and in
Reference [5].  It should be noted that none of the LEP experiments presently 
exploits the \WWlnln\ final state in the DR methods~\footnote{A measure of 
\Mw\ can be obtained from the \WWlnln\ channel by using the lepton energy 
spectrum.  However, it is estimated to be a factor of 4-5 less sensitive than 
the measurements available from the other \WW\ final states.}; it is therefore 
discussed no further.

\subsection{Event Selection}
\label{DR-evsel}

\noindent
The expected statistical error on \Mw\ varies as, $\Delta\Mw (\mrm{stat}) \sim 
    \frac{1}{\sqrt{\mrm{N_{WW}}}} \cdot \frac{1}{\sqrt{\mrm{Purity}}}$,
so that high efficiency, high purity selections are important.  The \WW\
selection efficiencies and purities are given in Table~\ref{tab:effpur} for
each of the four LEP experiments.

For the data taken at $\roots=183$~GeV, these efficiencies and purities give
approximately 700 \WW\ candidate events per experiment, about 100 of which are
non-\WW\ background.  The selection efficiencies have a total uncertainty 
of about $1\%$ (absolute) and have a negligible effect ($<1$~MeV) on the \Mw\ 
determination.  The accepted background cross-sections have a total uncertainty
of $20-40\%$ (relative) and effect the \Mw\ determination at the $10-15$~MeV 
level (cf. Section~\ref{sec:syserr}).
%
% --- selection efficiencies, purities
%
\begin{table}
  \caption{The \WW\ selection efficiency, $\varepsilon$, and purity, 
    $\mathcal{P}$, for the \qq\qq\ and \qq\lnu\ channels for each of
    the four LEP experiments.  \Delphi employs no explicit 
    \qq\tnu\ selection.}
  \vspace{4mm}
  \begin{center}
    \begin{tabular}{|lr|cccc|} \hline
      \multicolumn{2}{|c|}{channel} & \multicolumn{4}{c|}{experiment} \\
        & &A &D &L &O \\ \hline\hline
      $\qq\qq$ & $\varepsilon$ (\%)  & 83 & 85 & 88 & 85 \\
               & $\mathcal{P}$ (\%)  & 83 & 65 & 80 & 80 \\ \hline
      $\qq\enu(\mnu)$  & $\varepsilon$ (\%)  & 89 & 71 & 87 & 90 \\
                       & $\mathcal{P}$ (\%)  & 96 & 94 & 96 & 94 \\ \hline
      $\qq\tnu$ & $\varepsilon$ (\%) & 64 & -- & 59 & 75 \\
                & $\mathcal{P}$ (\%) & 93 & -- & 87 & 83 \\ \hline
    \end{tabular}
  \end{center}
  \label{tab:effpur}
\end{table}

\subsection{Invariant Mass Reconstruction}
\label{sec:mrec}

\noindent
There are several methods available for reconstructing the invariant mass of
a $\mrm{W}^{\pm}$ candidate.  The best resolution is obtained by using a 
kinematic fit which exploits the fact that the center-of-mass energy of the
collision is known {\it a priori}~\footnote{Strictly speaking, this is not true
since any initial state radiation (ISR) reduces the collision energy to less 
than twice the beam energy.  The kinematic fits assume no ISR.  The effect of
ISR uncertainties is incorporated in the total systematic error discussed in
Section~\ref{sec:syserr}.}.  The are two ``flavours'' of kinematic fit:
\begin{enumerate}
  \item 4C-fit: Enforces 
    $\Sigma(\mathbf{P},\mrm{E}) = (\mathbf{0},\roots)$ constraints; 
    yields {\it two} reconstructed masses per event, $(\mrecf,\mrecs)$, one 
    for each $\mrm{W}^{\pm}$ in the final state.
  \item 5C-fit: In addition to the four constraints above, ignores the finite
    width of the $\mrm{W}^{\pm}$ and requires that $\mrecf = \mrecs$;  yields
    a {\it single} reconstructed mass per event.
\end{enumerate}
The type of fit used depends on the final state.  For instance, in the \qq\enu\
and \qq\mnu\ channels, because the prompt neutrino from the leptonic
$\mrm{W}^{\pm}$ decay takes three degrees-of-freedom ($dof$), 
$\mathbf{P}_\nu$, the fits effectively become 1C and 2C fits 
respectively.  For the \qq\tnu\ channel, high energy neutrinos from the 
$\tau$-decay itself lose at least one additional $dof$ and so require that all
5 constraints be used, thus yielding a 1C fit~\footnote{Such a fit is possible 
only if one assumes that the $\tau$-lepton direction is given by the direction 
of the visible decay products associated with the $\tau$.}.  

In the
\qq\qq\ channel, since there are (nominally) four jets, there exist three
possible jet-jet pairings.  This pairing ambiguity gives rise to a combinatoric
background unique to the \qq\qq\ channel.  Each LEP experiment employs a
different technique for choosing the best combination(s).  \Opal and \Lt
use the 5C-fit probabilities (the equal mass constraint yields a different
fit $\chi^{2}$ for each combination) to choose the {\it two} best combinations
per event.  At the cost of some additional combinatorics, this algorithm
has the correct combination among those chosen about $90\%$ of the time.
\Delphi and \Aleph employ a 4C-fit and exploit information from the mass 
difference, $\Delta\mrec = \left| \mrecf - \mrecs \right|$, and the dijet 
opening angles to choose the best combination.  The algorithm employed by 
\Aleph chooses a single combination per event; this combination corresponds to
the correct combination approximately $85\%$ of the time at no additional cost
in combinatorics.  \Delphi uses all combinations and weights each according to 
the probability that it corresponds to the correct combination.

\subsection{Extracting $\mathbf{M_{W}}$}
\label{sec:extmw}

\noindent
The ensemble of selected events yields a \mrec\ distribution from which a 
measure of \Mw\ is extracted.  There are several methods available for 
extracting \Mw .  \Aleph, \Lt, and \Opal all employ a traditional maximum 
likelihood comparison of data to Monte Carlo (MC) spectra corresponding to
various \Mw.  In addition to its simplicity, this method has the advantage that
all biases (ie. from resolution, ISR, selection, etc.) are implicitly 
included in the MC spectra.  The disadvantage of this method is that it does
not make optimal use of all available information. \Delphi employs a 
convolution technique, which makes use of all available information;
in particular, events with large fit-errors are de-weighted relative to fits 
with small fit-errors.  The convolution has the limitations that 
it requires various approximations (ie. the resolution is often assumed to be
Gaussian) and often requires an {\it a posteriori} correction as the fit
procedure does not account for all biases, notably from ISR and selection.

%
% --- qqln results
%
\begin{table}
  \caption{ Results for data taken at $\roots=183$~GeV. All quantities are 
    given in units of GeV. }
  \vspace{4mm}
  \begin{center}
    \begin{tabular}{|c|c|c|} \hline
      \multicolumn{3}{|c|}{\qq\lnu\ channel} \\ \hline
      exp &$\Mw\pm(\mrm{stat})\pm(\mrm{syst})$
          & $\hat{\sigma}_{\mrm{stat}}$              \\ \hline\hline
      A & $80.16\pm0.20\pm0.08$ & $0.21$             \\
      D & $80.50\pm0.26\pm0.07$ & $0.25$             \\
      L & $80.03\pm0.24\pm0.07$ & $0.21$             \\
      O & $80.25\pm0.18\pm0.08$ & $0.20$             \\ \hline\hline
    LEP & $80.22\pm0.11\pm0.05$ & $\chi^{2} = 1.8/3$ \\ \hline 
    \end{tabular}
  \end{center}
  \label{tab:qqln183}
\end{table}
%
% --- qqqq results
%
\begin{table}
  \caption{ Results for data taken at $\roots=183$~GeV. All quantities are 
    given in units of GeV. }
  \vspace{4mm}
  \begin{center}
    \begin{tabular}{|c|c|c|} \hline
      \multicolumn{3}{|c|}{\qq\qq\ channel} \\ \hline
      exp &$\Mw\pm(\mrm{stat})\pm(\mrm{syst})\pm(\mrm{CR})$
          & $\hat{\sigma}_{\mrm{stat}}$           \\ \hline\hline
      A & $80.45\pm0.18\pm0.06\pm0.10$ & $0.18$   \\
      D & $80.02\pm0.20\pm0.05\pm0.10$ & $0.21$   \\
      L & $80.51\pm0.21\pm0.09\pm0.10$ & $0.22$   \\
      O & $80.48\pm0.23\pm0.08\pm0.10$ & $0.20$   \\ \hline\hline
    LEP & $80.36\pm0.10\pm0.05\pm0.10$ &$\chi^{2} = 3.8/3$ \\ \hline  
    \end{tabular}
  \end{center}
  \label{tab:qqqq183}
\end{table}

\section{Results}
\label{sec:res}

\begin{figure}
  \vspace*{-3mm}
  \begin{center}
    \epsfig{file=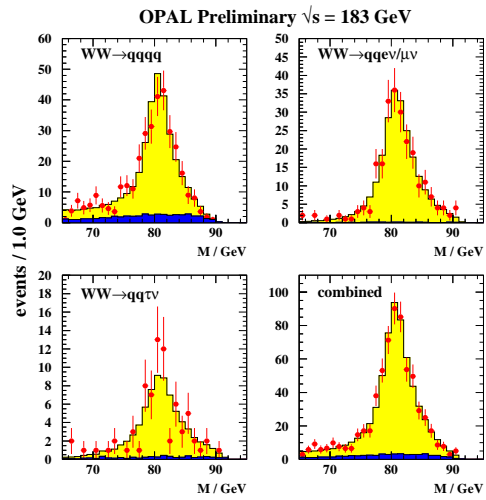,height=7cm}
  \vspace*{-2mm}
    \caption{Fit results for $\roots=183\:\mrm{GeV}$ data.  The points are 
      \Opal data, the histogram is the  fit result, and background 
      contributions are shown as the dark shaded regions.}
    \label{fig:ofit}
  \end{center}
\end{figure}

\noindent
The results from each LEP experiment, using data collected at $\roots=183$~GeV,
are given in Table~\ref{tab:qqln183} for \qq\lnu\ channel and in 
Table~\ref{tab:qqqq183} for the \qq\qq\ channel.  Also included is the mass 
obtained when combining all four measurements.  For the LEP combinations, the
ISR, hadronization, LEP beam energy, and color-reconnection/Bose-Einstein (CR)
uncertainties are taken as completely correlated between the four experiments.
{\it Note that these numbers are preliminary}.  The errors given correspond to
the observed statistical and the total systematic (including that associated 
with the LEP beam energy) uncertainties respectively.  For the \qq\qq\ channel,
the error associated with CR uncertainties is given separately and is taken as
a $100$~MeV common error. Also shown in 
Tables~\ref{tab:qqln183} and~\ref{tab:qqqq183} is the expected statistical 
error, $\hat{\sigma}_{\mrm{stat}}$, for each experiment.  As an example, 
the \Opal fits are shown in Figure~\ref{fig:ofit}.

Using all data taken at $\roots \geq 172$~GeV, the LEP combined \Mw\ using
DR methods for the \qq\lnu\ and \qq\qq\ channels separately is:
\bea
  \Mw (\qq\lnu) &= &80.27\pm0.10(\mrm{stat})\pm0.04(\mrm{syst})\:\mrm{GeV} \\  
  \Mw (\qq\qq)  &= 
      &80.40\pm0.09(\mrm{stat})\pm0.05(\mrm{syst})\pm0.10(\mrm{CR})\:\mrm{GeV}
\eea
These are consistent with one another within $1\sigma$ of their statistical 
errors.

\section{Systematic Errors}
\label{sec:syserr}

\noindent
The systematic errors for a typical LEP experiment are given in 
Table~\ref{tab:syserr}.  It should be noted that for all four LEP experiments
the errors associated with ISR, hadronization, and four-fermion interference
uncertainties are limited by the statistics of the comparison. Uncertainties
associated with the selection efficiencies and accepted backgrounds are 
included in the line labeled ``fit procedure''. For the \qq\lnu\ channel the 
single largest systematic uncertainty is due to uncertainties associated with 
detector effects (eg. energy scales, resolutions, and modelling).  These errors
are expected to decrease as more data is collected. For the \qq\qq\ channel the
dominant systematic uncertainty is due to CR effects.  An additional 
$50-100\:\mrm{pb}^{-1}$ of data will yield results that begin to constrain the
available models and thus improve these errors. See the summaries by G.Duckeck
and H.Hoorani in these proceedings for more details.

\section{Conclusions}
\label{sec:con}

\begin{figure}
  \begin{center}
    \epsfig{file=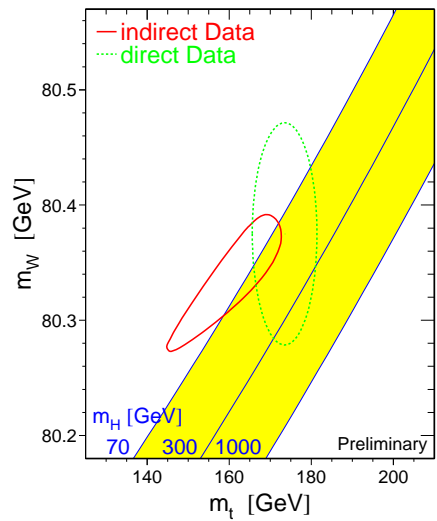,height=7cm}
    \vspace*{-2mm}
    \caption{Comparison of direct measurements of \Mw\ and $\mrm{M_{t}}$ to
      indirect measurements inferred from fits to precision electroweak data
      assuming the SM.}
    \label{fig:mwvmt}
  \end{center}
\end{figure}

\noindent
Using approximately $10\:\mrm{pb}^{-1}$ of data collected at 
$\roots=161$~and~172~GeV and $55\:\mrm{pb}^{-1}$ at $\roots=183$~GeV the LEP
experiments have measured the mass of the W boson.  The LEP combined result, 
including data from all center-of-mass energies and assuming the Standard Model
relation between the W decay width and mass, is 
$\Mw = 80.35 \pm 0.07 ( \mrm{exp} ) \pm 0.04 ( \mrm{CR} ) \pm 0.03 
  ( \mrm{E_{bm}} )$~GeV, 
where the errors correspond to experimental, colour-reconnection/Bose-Einstein,
and LEP beam energy uncertainties respectively. Figure~\ref{fig:mwvmt} compares
the $1\sigma$ contour for the direct measurements of 
\Mw~\cite{lepewk,tevmw} and $\mrm{M_{t}}$~\cite{tevmt} to the $1\sigma$ 
contour inferred from precision electroweak measurements at LEP, SLD, and the
TeVatron~\cite{lepewk}.  

During 1998 LEP is expected to deliver $150\:\mrm{pb}^{-1}$ per experiment at
$\roots\approx 190$~GeV.  This additional data will increase the present 
statistics available for the DR method by more than a factor of two and will
allow for experimental constraints on various color-reconnection and 
Bose-Einstein models in the \qq\qq\ final state.

\section*{Acknowledgements}

\noindent
Many thanks to my colleagues in the LEP Electroweak working group for their
comments and suggestions.

%
% --- systematic errors
%
\begin{table}
  \caption{ Table of systematic errors on \Mw\ for a typical LEP experiment. }
  \vspace{4mm}
  \begin{center}
    \begin{tabular}{|l|cc|} \hline
      systematic &\multicolumn{2}{c|}{ $\Delta\Mw$ (MeV)} \\
      source     & \qq\lnu & \qq\qq \\ \hline \hline
      initial state radiation &  15 &  15  \\ 
      hadronization           &  25 &  30  \\
      four fermion            &  20 &  20  \\
      detector effects        &  45 &  50  \\
      fit procedure           &  30 &  30  \\
      Sub-total               &  65 &  70  \\ \hline  
      beam energy             &  30 &  30  \\
      CR/BE                   &  -- &  100 \\ \hline\hline
      Total                   &  71 &  126 \\ \hline
    \end{tabular}
  \end{center}
  \label{tab:syserr}
\end{table}
%
% --- Opal fits figure
%
%
% --- Mw vs Mt
%

\end{document}